\documentclass[10pt,twocolumn]{IEEEtran}
\makeatletter
\def\subsubsection{\@startsection{subsubsection}
                                 {3}
                                 {\z@}
                                 {0ex plus 0.1ex minus 0.1ex}
                                 {0ex}
                             {\normalfont\normalsize\bfseries}}
\makeatother
\usepackage[T1]{fontenc}
\usepackage{subfigure}
\usepackage{ulem}
\usepackage{amsmath,enumitem}
\allowdisplaybreaks
\usepackage{hhline}
\usepackage{graphicx}
\usepackage{yfonts,color}
\usepackage{soul,xcolor}
\usepackage{verbatim}
\usepackage{flushend}
\usepackage{amsmath}
\allowdisplaybreaks
\usepackage{amssymb}
\usepackage{amsthm}
\usepackage{float}
\usepackage{bm}
\usepackage{url}
\usepackage{array}
\usepackage{cite}
\usepackage{tikz}
\usepackage{framed}
\usepackage{balance}
\usepackage{epsfig,epstopdf}
\usepackage{booktabs}
\usepackage{courier}
\usepackage{subfigure}
\usepackage{pseudocode}
\usepackage{enumerate}
\usepackage{algorithm}
\usepackage{algpseudocode}
\usepackage[binary-units]{siunitx}
\newcommand{\rom}[1]{\uppercase\expandafter{\romannumeral #1\relax}}
\usepackage{color}
\usepackage{soul,xcolor}

\usepackage{cancel}

\usepackage{setspace}
\usepackage{glossaries}
\newacronym{tx}{Tx}{transmitter}
\newacronym{rx}{Rx}{receiver}

\title{A Robotic Antenna Alignment and Tracking System for Millimeter Wave Propagation Modeling}
\author{\vspace{-1mm}Bharath Keshavamurthy\IEEEauthorrefmark{1}, Yaguang Zhang\IEEEauthorrefmark{2}, Christopher R. Anderson\IEEEauthorrefmark{3},\\Nicol\`{o} Michelusi\IEEEauthorrefmark{1}, James V. Krogmeier\IEEEauthorrefmark{2}, David J. Love\IEEEauthorrefmark{2}
\thanks{Part of this research has been funded by NSF under grant CNS-1642982.}
\thanks{\IEEEauthorrefmark{1}Electrical, Computer and Energy Engineering, Arizona State University.}
\thanks{\IEEEauthorrefmark{2}Electrical and Computer Engineering, Purdue University.}
\thanks{\IEEEauthorrefmark{3}Electrical Engineering, United States Naval Academy.}
\vspace{-12mm}}
\begin{document}
\bstctlcite{IEEEexample:BSTcontrol}
\maketitle
\thispagestyle{empty}
\pagestyle{empty}
\setulcolor{red}
\setul{red}{2pt}
\setstcolor{red}

\newcommand{\linespreadexceptabstractandindex}{\setstretch{0.98}} 
\newcommand{\linespreadforabstractandindex}{\setstretch{0.991}}

\newcommand{\extraspacebeforesec}{-4mm}
\newcommand{\extraspacebeforesubsubsec}{0.2mm}

\linespreadexceptabstractandindex

\begin{abstract}
\linespreadforabstractandindex
In this paper, we discuss the design of a sliding-correlator channel sounder for \SI{28}{\giga\hertz} propagation modeling on the NSF POWDER testbed in Salt Lake City, UT. Beam-alignment is mechanically achieved via a fully autonomous robotic antenna tracking platform, designed using commercial off-the-shelf components. Equipped with an Apache Zookeeper/Kafka managed fault-tolerant publish-subscribe framework, we demonstrate tracking response times of \SI{27.8}{\milli\second}, in addition to superior scalability over state-of-the-art mechanical beam-steering systems. Enhanced with real-time kinematic correction streams, our geo-positioning subsystem achieves a 3D accuracy of \SI{17}{\centi\meter}, while our principal axes positioning subsystem achieves an average accuracy of \SI{1.1}{\degree} across yaw and pitch movements. Finally, by facilitating remote orchestration (via managed containers), uninhibited rotation (via encapsulation), and real-time positioning visualization (via Dash/MapBox), we exhibit a proven prototype well-suited for V2X measurements.
\end{abstract}

\glsresetall

\vspace{\extraspacebeforesec}
\section{Introduction}\label{I}
With the widespread deployment of 5G networks by wireless carriers, primarily leveraging the mid-band spectrum, these service providers have shifted their spectrum procurement focus to the millimeter-wave bands (mmWave: \SIrange{30}{300}{\giga\hertz})~\cite{WSJ:Verizon}, with the long-term vision of providing a significant enhancement in user experience vis-\`{a}-vis data rates and latencies in dense urban and suburban environments. Concertedly, academic research into mmWave signal propagation modeling has also gained a renewed emphasis. In this paper, we briefly summarize our efforts in executing a measurement campaign with a \SI{28}{\giga\hertz} sliding-correlator channel sounder and a fully autonomous robotic antenna alignment \& tracking platform, on the NSF POWDER testbed at the University of Utah in Salt Lake City, UT~\cite{POWDER}. 

An earlier measurement campaign~\cite{Purdue} by our research group centered around a manual antenna alignment \& tracking platform in semi-stationary settings for \SI{28}{\giga\hertz} systems in suburban neighborhoods
. Similarly, the system detailed in~\cite{Harvard} also involves manual alignment and is restricted to indoor environments. On the other hand, the beam-alignment framework outlined here is fully autonomous and capable of operating in V$2$X mobility scenarios, with remote monitoring and troubleshooting capabilities; furthermore, this measurement campaign on the POWDER testbed is executed on a geographically diverse site encompassing both urban and suburban routes. Although electronic beam-alignment strategies involving phased-arrays offer faster switching times (${\approx}\SI{2.5}{\milli\second}$)~\cite{Agile-Link} and greater flexibility (side-lobe \& beam-width control) relative to mechanical fixed-beam steering, they constitute computationally expensive signal sampling along multiple directions and their design necessitates complex hardware due to resource-heavy algorithmic models. In the the following section, we briefly summarize the design of our prototype and demonstrate its efficacy through field-tested performance metrics.

\newcommand{\txrxfigwidth}{0.95\linewidth}
\begin{figure}[t]
    \begin{tabular}{cc}
        \begin{minipage}{0.49\linewidth} 
        	 \centering
            \includegraphics[width=\txrxfigwidth]{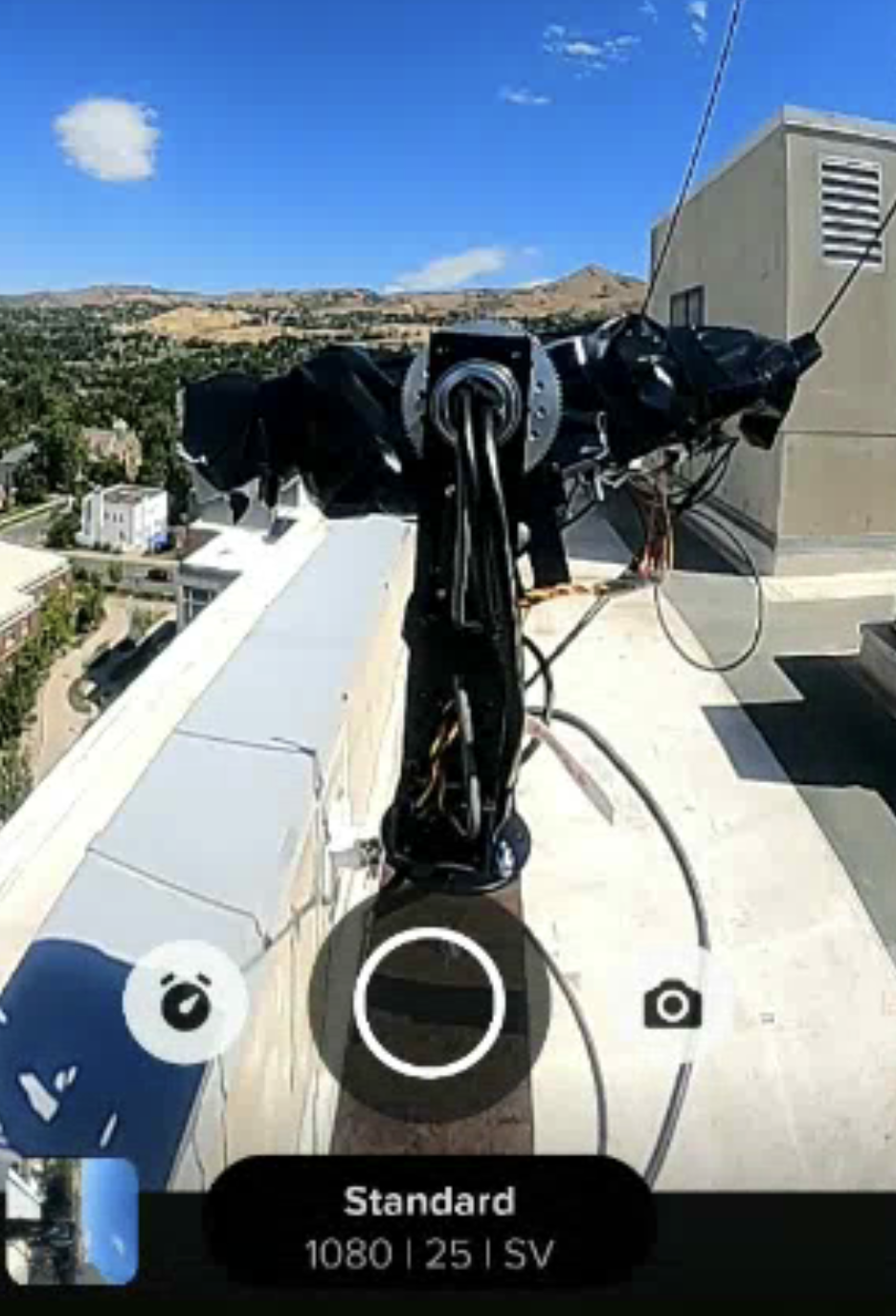}
            \\ [0.5ex]
            \centering
            \includegraphics[width=\txrxfigwidth]{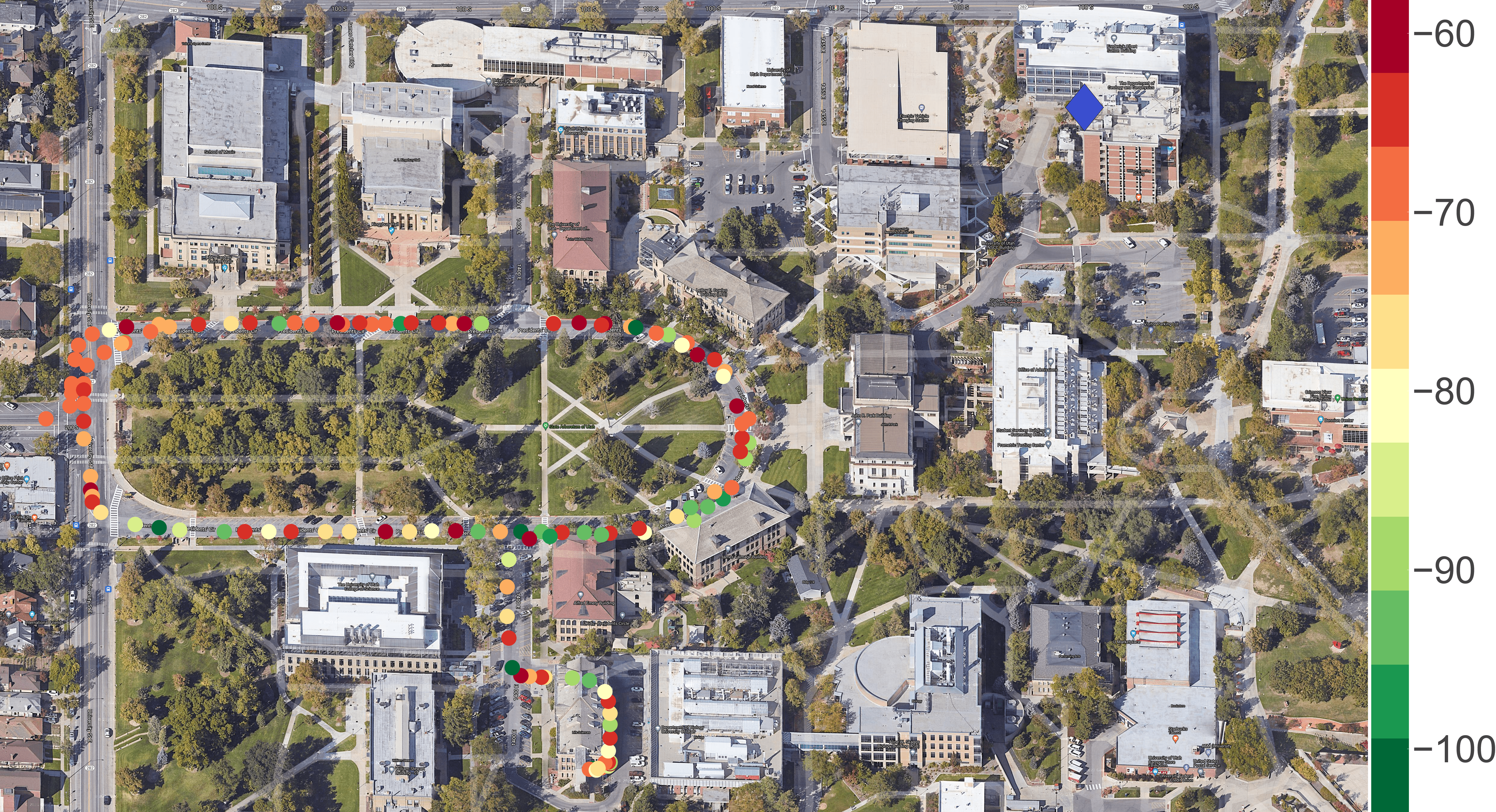}
        \end{minipage}&
        \hspace{-6mm}
        \begin{minipage}{0.51\linewidth}
        	 \centering
            \includegraphics[width=\txrxfigwidth]{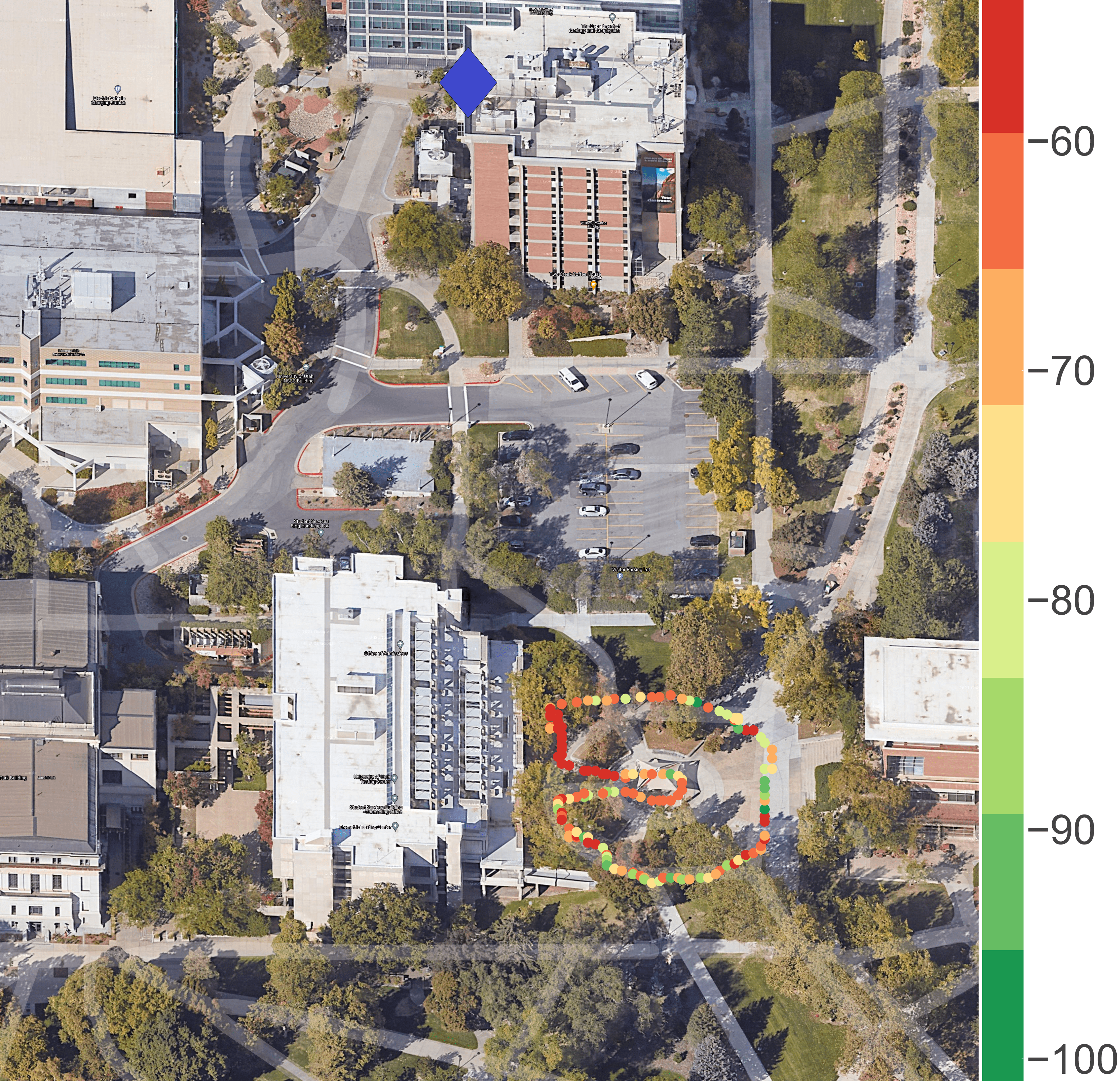}
            \\ [0.5ex]
            \centering
            \includegraphics[width=\txrxfigwidth]{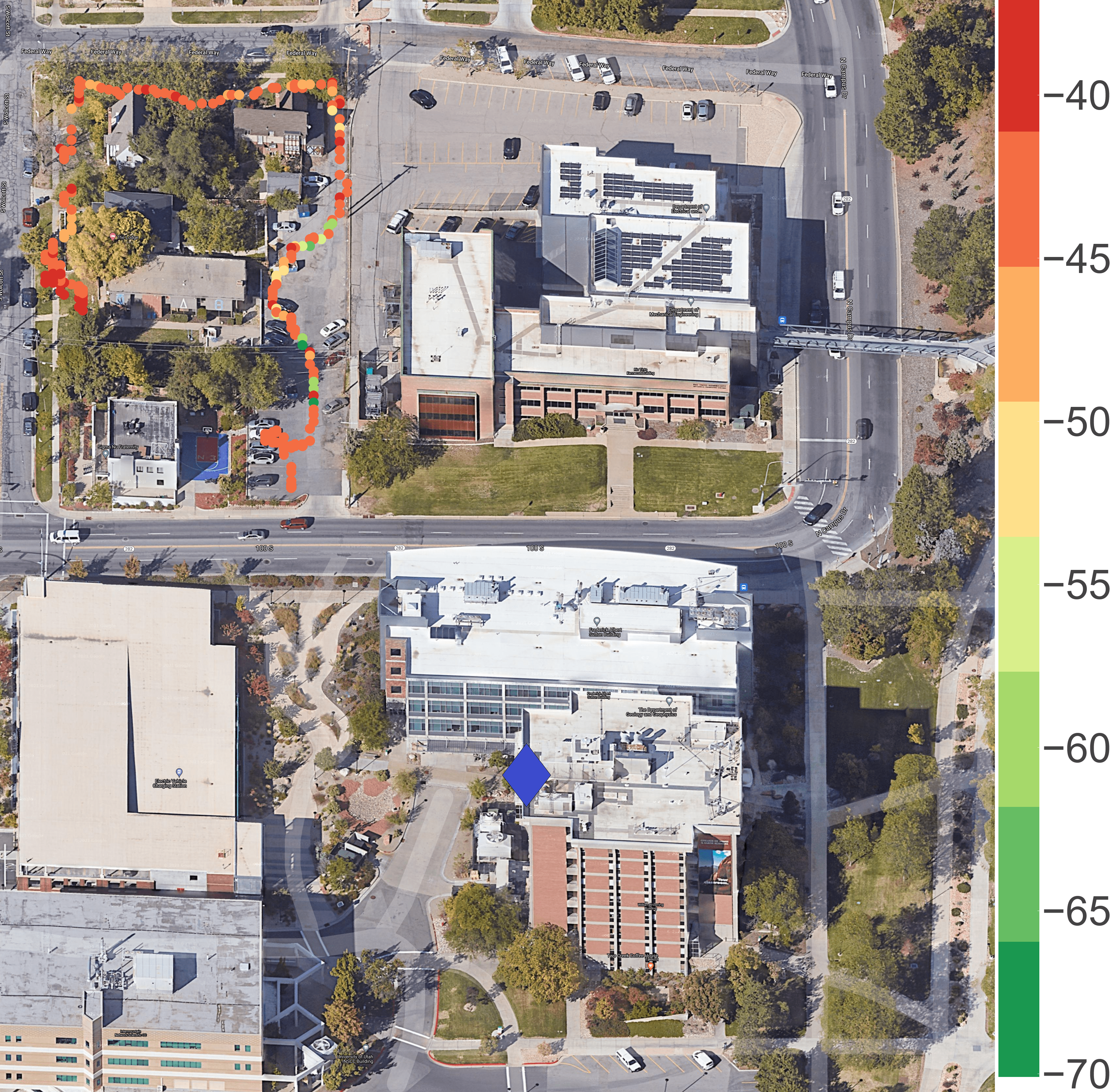}
        \end{minipage}
    \end{tabular}
    \caption{Clockwise from top-left: our remote monitoring \& troubleshooting interface---via an Android Debug Bridge---exhibiting the deployment of our \gls{tx} on the roof-top of the William Browning Building; illustrations of the received signal power values superimposed on a Google Hybrid map of the sites under analysis for urban foliage (Rx on cart), suburban neighborhood (Rx on cart), and urban-campus (Rx on minivan), respectively. The dots with heat-map color palette values denote \gls{rx} locations as it was driven/pushed around, and the purple diamond denotes the fixed \gls{tx} location.}
    \label{fig:TxRx}
    \vspace{-6mm}
\end{figure}

\begin{figure*} [t]
    \centerline{
    \includegraphics[width=1.0\textwidth]{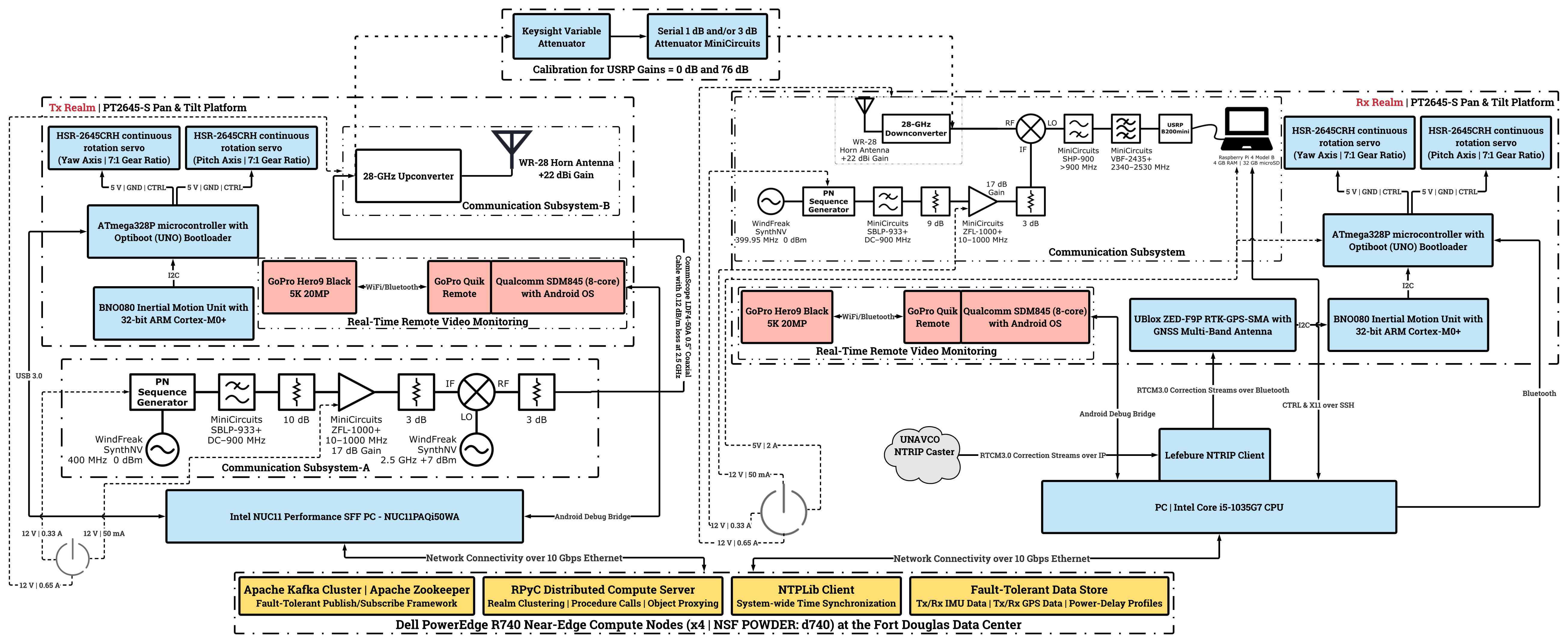}}
    \vspace{-4mm}
    \caption{The system architecture of our fully autonomous robotic antenna alignment \& tracking platform with a sliding-correlator channel sounder}
    \label{fig: Arch}
    \vspace{-6mm}
\end{figure*}

\vspace{\extraspacebeforesec}
\section{Design Description}\label{II}
The prototype deployed for our \SI{28}{\giga\hertz} measurement campaign on the POWDER testbed constitutes three modules: an autonomous \emph{beam-steering controller} replicated at both the \glsfirst{tx} and the \glsfirst{rx}, their respective \emph{communication subsystems} consistent with our sounder design, and a \emph{centralized nerve center} for component registration \& coordination, timing synchronization, messaging middleware, and data storage with redundancy. Here, the nerve center is deployed on a high-availability cluster of four Dell R740 compute nodes at the Fort Douglas data center; additionally, as depicted in Fig. \ref{fig:TxRx}, the \gls{tx} is fixed atop the William Browning Building, and the \gls{rx} is mounted on a Toyota Sienna minivan or a cart that is driven or pushed around campus.

\vspace{\extraspacebeforesubsubsec}\noindent{\bf Channel Sounder:} As illustrated in Fig. \ref{fig: Arch}, our prototype consists of a custom-built sliding-correlator channel sounder~\cite{Purdue}, whose principles have been laid down in~\cite{Sounder}. Using highly directional WR-$28$ horn antennas with $+$\SI{22}{\deci\bel{i}} gain \& \SI{15}{\degree} half-power beam-width, up/down-converters, and commercial off-the-shelf circuitry, we record the power-delay profiles (as complex-$64$ I/Q samples) at a Raspberry Pi via a GNURadio sink coupled with a USRP B$200$mini UHD source (bandwidth${=}$\SI{0}{\hertz}, gain${=}$\num{0} or \SI{76}{\deci\bel}, center frequency${=}$\SI{2.5}{\giga\hertz}, and sample rate${=}$\SI{2}{\mega{sps}}). Furthermore, leveraging QT GUI time sink visualization capabilities of GNURadio (with dynamic trigger levels) over an ad-hoc WLAN with the Pi, this design enables seamless oversight of our recording activities.

\vspace{\extraspacebeforesubsubsec}\noindent{\bf Alignment \& Tracking:} As shown in Fig. \ref{fig: Arch}, our mechanical beam-steering controller (written in C/C++ and Python) is inherited and extended at the \gls{tx} and the \gls{rx}. Each side interfaces with inertial motion and global positioning breakout boards through an ATMega328P microcontroller over I2C peripherals, gets notified of alignment \& location updates from its counterpart in real-time, and orchestrates yaw-/pitch-axis positioning via open-loop servo (HSR-2645CRH) control. The RPyC API handles controller registration \& coordination. A Network Time Protocol client-server architecture administers timing synchronization. A Zookeeper-managed cluster of Kafka brokers handles the publish-subscribe middleware operations between the \gls{tx} and \gls{rx} controllers, with redundant data storage. This fully autonomous architecture allows for an average beam-steering response time of \SI{27.8}{\milli\second}, evaluated over $12,870$ interactions; moreover, our loosely coupled modular design allows for remote monitoring and troubleshooting of each controller. Corroborated both in-field and on an ad-hoc laser testbed, our principal axes positioning framework achieves an accuracy of \SI{1.1}{\degree} across all fine- \& coarse-grained yaw and pitch movements. Likewise, corroborated both in-field (Plotly Dash/MapBox API) and on a vendor-specific console (uBlox u-Center/Google Maps JavaScript API), our geo-location framework---with real-time kinematic correction streams---achieves a $3$D accuracy of \SI{17}{\centi\meter}.

\vspace{\extraspacebeforesubsubsec}\noindent{\bf Post-Processing:} Upon completion of our data collection activities on the POWDER testbed, for a chosen site, our post-processing scripts parse the metadata associated the recorded power-delay profiles, extract the timestamps corresponding to each recorded data segment ($10^6$ complex-$64$ I/Q samples), process the samples in each segment (pre-filtering, temporal truncation, time-windowing, noise elimination), compute the received signal power off of these processed segments (with an initial calibration driven correction), couple these segments with the recorded geo-location \& inertial motion logs, and visualize the results on a Google Hybrid map rendered via the Bokeh toolbox: these visualizations for a \gls{rx} route around foliage in an urban campus environment, a suburban neighborhood, and President's Circle are shown in Fig. \ref{fig:TxRx}.

\vspace{\extraspacebeforesec}
\section{Conclusion}\label{III}
In this summary, we outline the key design details encompassing our mmWave propagation modeling measurement prototype. With several sites successfully analyzed at the University of Utah in Salt Lake City, our system has been proven both in the field and in a laboratory environment to achieve an average alignment accuracy of \SI{1.1}{\degree}, a $3$D geo-location accuracy of \SI{17}{\centi\meter}, and an average beam-alignment response time of \SI{27.8}{\milli\second} under V$2$X mobility evaluations. Offering unrestrained rotation and remote orchestration capabilities, in addition to a modular fault-tolerant messaging middleware framework, our fully autonomous beam-steering controller is well-suited to be scaled to increasingly complex mmWave modeling activities.

\vspace{0mm}

\balance
\bibliographystyle{IEEEtran}
\vspace{-4mm}
\bibliography{IEEEabrv,main}

\begin{thebibliography}{1}
\providecommand{\url}[1]{#1}
\csname url@samestyle\endcsname
\providecommand{\newblock}{\relax}
\providecommand{\bibinfo}[2]{#2}
\providecommand{\BIBentrySTDinterwordspacing}{\spaceskip=0pt\relax}
\providecommand{\BIBentryALTinterwordstretchfactor}{4}
\providecommand{\BIBentryALTinterwordspacing}{\spaceskip=\fontdimen2\font plus
\BIBentryALTinterwordstretchfactor\fontdimen3\font minus
  \fontdimen4\font\relax}
\providecommand{\BIBforeignlanguage}[2]{{%
\expandafter\ifx\csname l@#1\endcsname\relax
\typeout{** WARNING: IEEEtran.bst: No hyphenation pattern has been}%
\typeout{** loaded for the language `#1'. Using the pattern for}%
\typeout{** the default language instead.}%
\else
\language=\csname l@#1\endcsname
\fi
#2}}
\providecommand{\BIBdecl}{\relax}
\BIBdecl

\bibitem{WSJ:Verizon}
\BIBentryALTinterwordspacing
S.~Krouse, ``{U.S.} telecom giants take different paths to {5G},'' \emph{The
  Wall Street Journal}, 2020. [Online]. Available:
  \url{https://www.wsj.com/articles/u-s-telecom-giants-take-different-paths-to-5g-11586556074}
\BIBentrySTDinterwordspacing

\bibitem{POWDER}
J.~Breen \emph{et~al.}, ``{POWDER}: Platform for open wireless data-driven
  experimental research,'' in \emph{Proceedings of the 14th International
  Workshop on Wireless Network Testbeds, Experimental Evaluation and
  Characterization (WiNTECH)}, Sep. 2020, doi: \url{10.1145/3411276.3412204}.

\bibitem{Purdue}
Y.~Zhang \emph{et~al.}, ``{28-GHz} channel measurements and modeling for
  suburban environments,'' in \emph{2018 IEEE Intl. Conf. Commun.}, 2018, pp.
  1--6.

\bibitem{Harvard}
M.~Comiter \emph{et~al.}, ``Millimeter-wave field experiments with many antenna
  configurations for indoor multipath environments,'' in \emph{2017 IEEE
  Globecom Workshops (GC Wkshps)}, 2017, pp. 1--6.

\bibitem{Agile-Link}
H.~Hassanieh \emph{et~al.}, ``Fast millimeter wave beam alignment,'' in
  \emph{2018 ACM Special Interest Group on Data Communication (SIGCOMM)}.\hskip
  1em plus 0.5em minus 0.4em\relax New York, NY, USA: Association for Computing
  Machinery, 2018, p. 432–445, doi: \url{10.1145/3230543.3230581}.

\bibitem{Sounder}
R.~J. Pirkl and G.~D. Durgin, ``Optimal sliding correlator channel sounder
  design,'' \emph{{IEEE} Trans. Wireless Commun.}, vol.~7, no.~9, pp.
  3488--3497, 2008.

\end{thebibliography}

\end{document}